\documentclass[
reprint,
superscriptaddress,
 amsmath,amssymb,
 aps,longbibliography,
pra,
floatfix,
toc=flat,
]{revtex4-1}

\usepackage{comment}
\usepackage{graphicx}
\usepackage{txfonts}
\usepackage{hyperref}
\usepackage[usenames,dvipsnames]{xcolor}
\usepackage{braket}
\usepackage{ bbold }
\usepackage{placeins}
\usepackage{lipsum}
\usepackage[utf8x]{inputenc}
\usepackage{color}
\usepackage{bbm} 
\usepackage{amsfonts,amsmath,amssymb,stmaryrd}
\usepackage{subfigure}
\usepackage{bbm} 
\usepackage{epsfig}
\usepackage{mathrsfs}
\usepackage{verbatim}
\usepackage{centernot}
\usepackage{nicefrac}
\DeclareMathOperator{\Tr}{Tr}

\definecolor{blue(pigment)}{rgb}{0.2, 0.2, 0.6}
\definecolor{darkerblue}{rgb}{0.0, 0.0, 0.4}
\definecolor{darkblue}{rgb}{0.0,0.0,0.5}
\definecolor{darkgreen}{rgb}{0.0,0.4,0.0}
\hypersetup{
   colorlinks,
   linkcolor=blue(pigment),
    citecolor=darkgreen,
   urlcolor=darkblue
}

\begin{document}

\title{Probing finite-temperature observables in quantum simulators of spin systems\\ with short-time dynamics}

\begin{abstract}
Preparing finite temperature states in quantum simulators of spin systems, such as trapped ions or Rydberg atoms in optical tweezers, is challenging due to their almost perfect isolation from the environment.  Here, we show how finite-temperature observables can be obtained with an algorithm motivated from the Jarzynski equality and equivalent to the one in \href{https://doi.org/10.1103/PRXQuantum.2.020321}{Lu, Ba\~nuls and Cirac, PRX Quantum 2, 020321 (2021)}. It consists of classical importance sampling of initial states and a measurement of the Loschmidt echo with a quantum simulator. We use the method as a quantum-inspired classical algorithm and simulate the protocol with matrix product states to analyze the requirements on a quantum simulator. This way, we show that a finite temperature phase transition in the long-range transverse field Ising model can be characterized in trapped ion quantum simulators. We propose a concrete measurement protocol for the Loschmidt echo and discuss the influence of measurement noise, dephasing, as well as state preparation and measurement errors. We argue that the algorithm is robust against those imperfections under realistic conditions. 
\end{abstract}

\author{Alexander Schuckert}
\affiliation{Department of Physics, Technical University of Munich, 85748 Garching, Germany}
\affiliation{Munich Center for Quantum Science and Technology (MCQST), 80799 M\"unchen, Germany}
\affiliation{Quantinuum, Leopoldstrasse 180, 80804 Munich, Germany}
\author{Annabelle Bohrdt}
\affiliation{ITAMP, Harvard-Smithsonian Center for Astrophysics, Cambridge, MA 02138, USA}
\affiliation{Department of Physics, Harvard University, Cambridge, Massachusetts 02138, USA}
\author{Eleanor Crane}
\affiliation{Department of Electrical Engineering and London Centre for Nanotechnology, University College London, Gower Street, London WC1E 6BT, United Kingdom}
\affiliation{Quantinuum, Leopoldstrasse 180, 80804 Munich, Germany}
\author{Michael Knap}
\affiliation{Department of Physics, Technical University of Munich, 85748 Garching, Germany}
\affiliation{Munich Center for Quantum Science and Technology (MCQST), 80799 M\"unchen, Germany}
\date{\today}
\maketitle

The excellent tunability of quantum simulators has enabled new insights into entire classes of many-body models.
Recently, tremendous progress has been achieved in simulating unconventional nonequilibrium dynamics of quantum spin models with a large number of controlled degrees of freedom on different experimental platforms, including ultracold atoms in optical lattices~\cite{Hild2014, Brown2015, Guardado-Sanchez2018, Jepsen2020, Wei2022}, Rydberg atoms~\cite{Labuhn2016, Bernien2017, Pineiro2018}, and trapped ions~\cite{Britton2012,Zhang2017_2, Joshi2022}. A key question motivated from condensed matter physics is to study finite temperature states of quantum spin models, which can host phases with symmetry breaking or even topological order, thermal phase transitions, and quantum criticality~\cite{Sachdev2011}. Preparing states at finite and in particular low temperatures, required to study these phenomena, is however a formidable challenge for quantum simulators because of their almost perfect isolation from the environment. 

\begin{figure}[ht!]
\centering
\includegraphics[width=\linewidth]{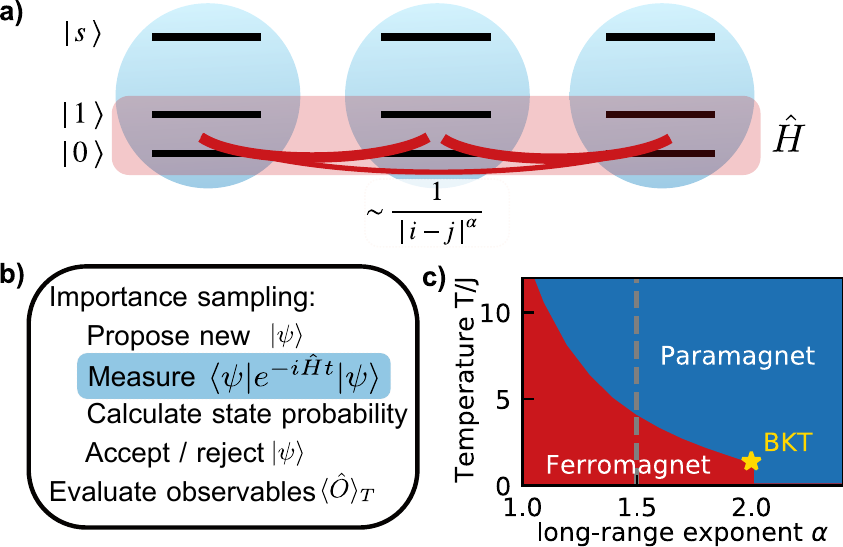}
\caption{\textbf{Setup and phase diagram.} a) Trapped ions in internal states $\ket{0}$, $\ket{1}$ interact with powerlaw couplings with exponent $\alpha$. Ions in the shelving state $\ket{s}$ do not interact. b) Thermal observables are obtained from Monte Carlo importance sampling of product states; see Ref.~\cite{Lu2021}. The probability of proposed states is classically evaluated from the Loschmidt echos $G_\psi(t)$ measured on the quantum simulator (blue shading), in which the shelving state $\ket{s}$ is used for obtaining the phase of $G_\psi(t)$ with Ramsey experiments. The Loschmidt echo only needs to be measured to short-times $Jt=\mathcal{O}(1)$~\cite{Lu2021}. c) The one-dimensional long-range transverse field Ising model exhibits a finite temperature phase transition from a ferromagnet to a paramagnet for $\alpha\leq 2$ (and transverse field $g <  g_c$; here $g=J$). At $\alpha=2$, the transition is in the Berenzinskii-Kosterlitz-Thouless universality class. In this work, we focus on $\alpha=1.5$ (grey dashed line).} 
 \label{fig:1}
 \end{figure}

Here, we discuss a concrete approach how to measure finite temperature observables in quantum simulation experiments based on an algorithm recently derived by Lu, Ba\~nuls and Cirac~\cite{Lu2021}. We motivate this algorithm from a different vantage point based on the Jaryznski equality~\cite{Jarzynski1997}, which provides a link between the nonequilibrium dynamics of a quantum system and its thermal properties. The key result of this algorithm is to obtain thermal observables from quantum simulators without preparing a thermal state directly, but to use a short real-time evolution instead. While this algorithm can be immediately applied to current quantum technology, even for large systems, other algorithms proposed for preparing finite temperature states, including sampling methods~\cite{Poulin2009, Temme2011, Bilgin2010, Chowdhury2017,Cohn2020}, imaginary time evolution~\cite{Motta2020, Sun2021}, variational methods~\cite{Zhu2020}, kernel based methods~\cite{Wang2022} and direct state preparation using fluctuation theorems~\cite{Holmes2022} are challenging to implement on devices without fault-tolerance due to significant resource overheads. Contrarily, the algorithm from Ref.~\cite{Lu2021} can be implemented on current devices due to its low requirement on evolution time and its error resilience.

We apply this algorithm to the detection of the thermal phase transition in the one-dimensional long-range transverse field Ising model (LTFIM)
\begin{equation}
    \hat H= - J\sum_{i<j} \frac{1}{|i-j|^\alpha} \hat \sigma^z_i\hat \sigma^z_j - g\sum_i \hat \sigma^x_i
    \label{eq:H_LTFI},
\end{equation}
where $g$ is the strength of the transverse field,  $J/|i-j|^\alpha>0$ is the ferromagnetic coupling with long-range exponent $\alpha$, and $\hat\sigma^a_i$ is the $a$-th Pauli matrix on site $i$. This model can be implemented with trapped ions~\cite{Porras2004}; see Fig.~\ref{fig:1}a), c). For $\alpha\leq 2$, the system exhibits a finite temperature transition from a ferromagnet at low temperatures to a paramagnet at high temperatures, provided the transverse field is sufficiently weak~\cite{Dutta2001, Knap2013}, c.f. also Ref.~\onlinecite{Islam2011} for an experiment with adiabatic state preparation. 

The Jarzynski-inspired algorithm uses Monte Carlo (MC) importance sampling of product states, where the state probabilities are classically calculated from the Loschmidt echo $G_\psi(t)=\braket{\psi|e^{-i\hat H t}|\psi}$ that is measured on a quantum simulator; Fig.~\ref{fig:1}b). We develop a scheme for measuring $G_\psi(t)$ based on Ramsey spectroscopy that involves a shelving state; Fig.~\ref{fig:1}a). The Loschmidt echo $G_\psi(t)$ only needs to be evaluated to short times, as proven in Ref.~\onlinecite{{Lu2021}}. To benchmark the algorithm, we interpret it as a quantum-inspired classical algorithm by evaluating the Loschmidt echos $G_\psi(t)$ with matrix product states (MPS). We find that the finite temperature phase transition of the LTFIM can be efficiently characterized with this algorithm even for large systems. To assess the feasibility of the algorithm in a realistic quantum simulator, we study its robustness to a finite number of measurements and discuss dephasing noise as well as state preparation and measurement (SPAM) errors, demonstrating the immediate applicability of the algorithm in current experimental technology. 

\emph{\textbf{Thermal properties from the Jarzynski equality.---}}The algorithm of Ref.~\cite{Lu2021} can be motivated from the Jarzynski equality~\cite{Jarzynski1997}, which is based on the following thought experiment~\cite{Talkner2007, Silva2008}: A system is prepared in thermal equilibrium at temperature $T$ with respect to a Hamiltonian $\hat H_0$. A measurement of $\hat H_0$ is then performed, which projects the system with probability $\frac{1}{Z_0}e^{-E_n^{0}/T}$ into the energy eigenstate $\ket{n^0}$ with energy $E_n^{0}$, where $Z_0$ is the partition sum of $\hat H^{0}$. Then, a second measurement in the eigenbasis of $\hat H$ is performed,  which yields the result $E_m$ with probability $|\braket{n^{0}|m}|^2$. In this process, the energy of the system changed and therefore work $\omega=E_m-E_n^{0}$ has been performed. Repeating this experiment many times, we can measure the probability distribution of work $\omega$. It is given by $p(\omega)=\frac{1}{Z_0}\sum_n e^{-E_n^{0}/T} \sum_m |\braket{n^{0}|m}|^2\delta(\omega-(E_m-E_n^{0}))$. By multiplying the work distribution with $e^{-\omega/T}$ and integrating, we find the Jarzynski equality
\begin{equation}
    Z=Z_0 \int d\omega e^{-\omega/T} p(\omega),
    \label{eq:Jarzynski}
\end{equation}
which relates an equilibrium quantity, the thermal partition sum $Z=\Tr(e^{-\hat H/T})$, to a non-equilibrium quantity, the work distribution. While in principle, all properties of a system can be obtained from $Z$, it is in practice hard to evaluate. Here we focus on the evaluation of finite temperature observables $\braket{\hat O}_T = \frac{1}{Z}\Tr\left(\hat O e^{-\hat H/T} \right)$.
In particular, by formally choosing $\hat H_0\propto \mathbb{1}$ 
we can use the Jarzynski equality even without preparing a thermal state of $\hat{H}_0$, since for this choice, the dependence on $\hat H_0$ becomes trivial. 

In order to relate the work distribution function to the Loschmidt echo, we interpret work as the Fourier conjugate to time by writing the Delta-function as $\delta(\omega)=\int\frac{dt}{2\pi} e^{i\omega t}$. Then, the Jarzynski equality becomes $Z=\int d\omega e^{-\omega/T} \int \frac{dt}{2\pi} e^{i\omega t} \Tr\left(e^{-i\hat H t}\right)$.
We expand the trace in a basis of product states $ \mathbb{1} = \sum_\psi \ket{\psi}\bra{\psi}$  to write $Z=\sum_\psi\int d\omega e^{-\omega/T}  p_\psi(\omega)$ with
\begin{equation}
   p_\psi(\omega)=\int \frac{dt}{2\pi} e^{i\omega t} \bra{\psi}e^{-i\hat H t}\ket{\psi}.\label{eq:ppsiomega}
\end{equation}
This way, we have reduced the evaluation of $Z$ via the Jarzynski equality to a measurement of the Loschmidt echo $G_\psi(t)=\bra{\psi}e^{-i\hat H t}\ket{\psi}$ with respect to product states $\ket{\psi}$ without requiring to prepare a thermal state in the quantum simulator.

\begin{figure}
\centering
\includegraphics{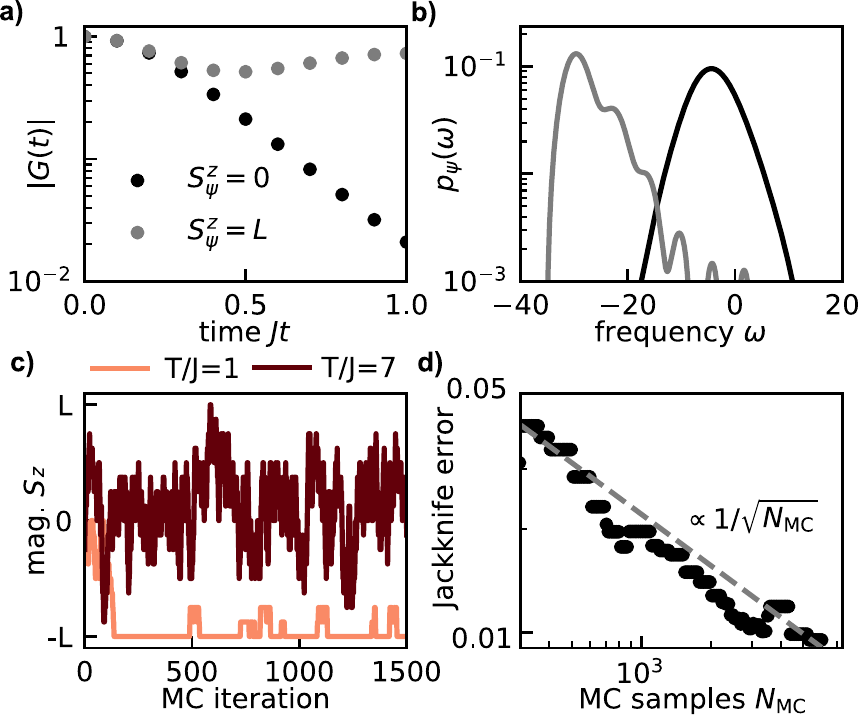}
\caption{\textbf{Data processing with classical resources.} a) Loschmidt echo shown for two different initial states---a completely polarized state and a state with vanishing magnetization. This data can be measured directly on a quantum simulator or can be obtained from numerical simulations for the quantum-inspired classical algorithm. b) The corresponding work distributions $p_\psi(\omega)$ are centred around the energies of the initial states, with the width given by the energy fluctuations. c) Magnetization evaluated from Monte Carlo importance sampling. d) The absolute error of the squared magnetization $\braket{(S^z)^2}$ at temperature $T/J=7$ calculated by a Jackknife estimate. 
Parameters: $L=16$, $\delta/J=4$, $Jt_\mathrm{max}=1$, $\Delta t=0.1$.}
 \label{fig:2}
 \end{figure}
 
In order to obtain observables $\hat O$ from $Z$, we shift $\hat H\rightarrow \hat H +h\hat O$ and evaluate 
\begin{equation}
    \braket{\hat O}_T = -\frac{T}{Z} \frac{dZ}{dh}\bigg|_{h=0}.
\end{equation}
While in principle any observable can be evaluated this way (see supplement~\cite{supp}), the simplest algorithm can be derived for observables satisfying $\hat O \ket{\psi} =O_\psi \ket{\psi}$. Inserting the Jarzynski equality, Eq.~\eqref{eq:Jarzynski}, we find in this case
\begin{equation}
   \braket{\hat O}_T = \frac{\sum_\psi  p_\psi(T) O_\psi}{\sum_\psi p_\psi(T)},
\label{eq:observables}
\end{equation}
where 
\begin{equation}
    p_\psi(T)=\int  p_\psi(\omega) e^{-\omega/T}d\omega.
    \label{eq:p_psi}
\end{equation}

Due to the exponential size of the Hilbert space, we cannot evaluate Eq.~\eqref{eq:observables} exactly. However, because $ p_\psi(T)/\sum_\psi p_\psi(T)$ is a probability distribution (even for frustrated and fermionic models), we can use a classical Monte Carlo importance sampling algorithm to select product initial states:\\
(1) Start with some product state $\ket{\psi}$.\\
(2) Repeat $N_\mathrm{MC}$ times for a given temperature $T$:\\
\hspace*{10pt} (a) Propose a new state $\ket{\psi'}$.\\
\hspace*{10pt} (b) Measure $G_{\psi'}(t)$ on a quantum simulator.\\
\hspace*{10pt} (c) Evaluate $p_{\psi'}(\omega)$ from Eq.~\eqref{eq:ppsiomega}.\\
\hspace*{10pt} (d) Evaluate $p_{\psi'}(T)$ from Eq.~\eqref{eq:p_psi}.\\
\hspace*{10pt} (e) Accept $\ket{\psi'}$ with probability $p_{\psi'}(T)/p_{\psi}(T)$.\\
\hspace*{10pt} (f) Evaluate $O_\psi$.\\
(3) Obtain thermal expectation value by averaging $O_\psi$.\\
This algorithm is the same as the one introduced in  Ref.~\onlinecite{Lu2021} based on energy filtering. Here, we provide an interpretation of this algorithm based on the Jarzynski equality. 

In the following, we discuss each step in more detail and apply it to the LTFIM. For step (1), a basis needs to be chosen. We use z-product states for which $\hat \sigma^z_i \ket{\psi}=\sigma^z_i \ket{\psi}$. As a state proposal in step (2a) we flip a single, randomly chosen spin, which fulfills ergodicity and, together with the acceptance step (2e), detailed balance~\cite{METROPOLIS1953}. 

In step (2b), we measure $G_{\psi}(t)$ on the quantum simulator in a time interval $[0,t_\mathrm{max}]$ with equal time steps $\Delta t$; Fig.~\ref{fig:2}a).  
The evaluation of $G_\psi(t)$ is the only step performed on the quantum simulator---all other steps use classical resources and take negligible computation time. We will discuss below how to measure this quantity in trapped ion simulators.

From $G_\psi(t)$, we then evaluate $p_{\psi}(\omega)$ from Eq.~\eqref{eq:ppsiomega} by a discrete Fourier transform in step (2c),
\begin{equation}
    p_{\psi}(\omega)=\frac{\Delta t}{2\pi} \sum_{n=-N}^{n=N} e^{i\omega n\Delta t}   G_{\psi}(t) e^{-\frac{(t \delta)^2}{2}},
    \label{eq:DFT}
\end{equation}where $\omega=2\pi n/(\Delta t (2N+1))$, $N=t_\mathrm{max}/\Delta t$. Above, we have introduced a Gaussian filter with standard deviation $1/\delta$ in order to suppress artifacts due to the finite $t_\mathrm{max}$; Fig.~\ref{fig:2}b).
Because $p_\psi(\omega)$ in Eq.~\eqref{eq:ppsiomega} is the density of states weighted by the overlap of $\ket{\psi}$ with the eigenstates, its width is given by the energy fluctuations $g\sqrt{L}$ (see supplement~\cite{supp}). Hence, in order for the frequency range of the discrete Fourier transform to cover $p(\omega)$, the timestep needs to be scaled as $\Delta t\propto 1/\sqrt{L}$. The width of the Gaussian filter $\delta$ can be chosen independently of $L$ and, hence, $t_\mathrm{max}\sim 1/\delta$. In fact, even a scaling $\delta \propto \sqrt{L}$ leads to convergence (with $t_\mathrm{max}\sim 1/\sqrt{L}$) leaving the number of time step evaluations constant~\cite{Lu2021, Yang2022}. 
 
From $p_\psi(\omega)$, we classically evaluate $p_{\psi}(T)$ according to Eq.~\eqref{eq:p_psi} in step (2e). In this evaluation, small errors introduced by experimental errors or numerical imprecision are exponentially amplified at low temperatures for large negative $\omega$ and low $T$. To mitigate this problem, we set $p_{\psi}(\omega)=0$ when $p_{\psi}(\omega)<p_\mathrm{cut}$. Moreover, $p_\psi(T)$ is centred around $E_\psi= \braket{\psi|\hat H|\psi}$, which can be problematic as $E_\psi\propto L$ if $\psi$ has a large overlap with states on the edges of the spectrum. This is for example the case for the totally polarized state in Fig.~\ref{fig:2}a). In order to resolve the fast oscillation, $\Delta t\propto 1/L$ would have to be chosen. However, we can circumvent this problem by shifting the zero of the frequency by $E_\psi$ when evaluating Eqs.~\eqref{eq:DFT} and \eqref{eq:p_psi}. This guarantees that $\Delta t\propto 1/\sqrt{L}$.

Having evaluated $p_\psi(T)$, we can now accept the state with probability $p_{\psi'}(T)/p_\psi(T)$ (step 2e) and store the value of $O_\psi$ of the state after the acceptance step; Fig.~\ref{fig:2}c). After repeating the importance sampling iteration for $N_\mathrm{MC}$ times, we evaluate thermal observables by averaging the $O_\psi$ (step (3)). For the LTFIM, we evaluate the squared magnetization $(\hat S^z/L)^2$ as well as the Binder cumulant $(3/2)-\braket{(\hat S^z)^4}/(2\braket{(\hat S^z)^2}^2)$ by calculating a power of the magnetization $\hat S^z=\sum_i \hat \sigma^z_i$ in the importance sampling. The Binder cumulant is a standard observable for the detection of Ising phase transitions as the Binder cumulant approaches $1\,(0)$ in the ferromagnetic (paramagnetic) phase~\cite{Binder1981}. Because importance sampling creates correlated samples, we use a  Jackknife binning analysis to determine error bars. In Fig.~\ref{fig:2}d) we show that these errors scale as $1/\sqrt{N_\mathrm{MC}}$ as expected from the central limit theorem. 

 \begin{figure}
\centering
\includegraphics[width=\linewidth]{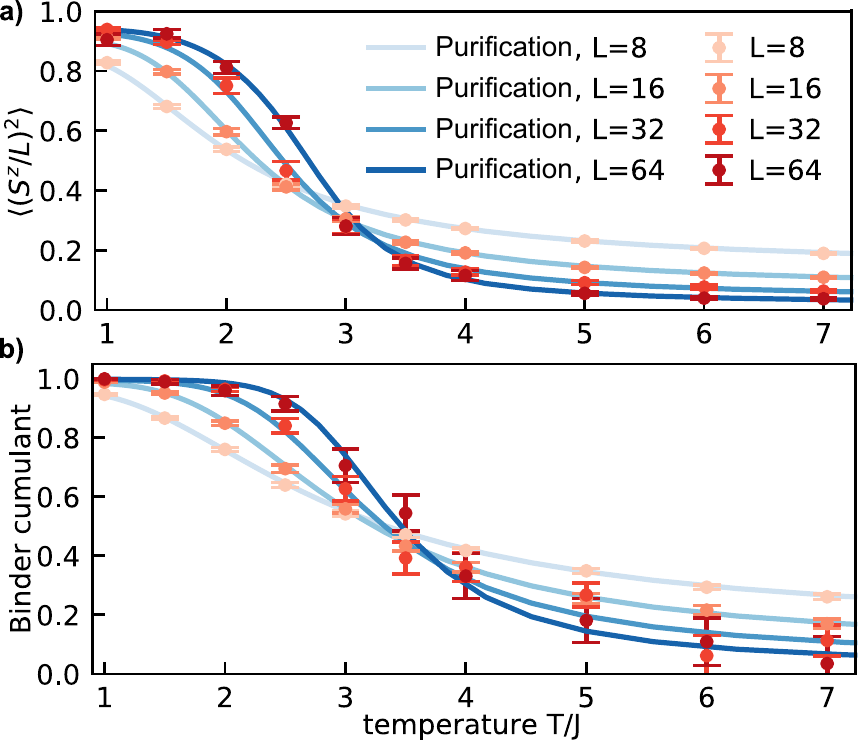}
\caption{\textbf{Detecting the phase transition in the LTFIM}. Data points for a) the squared magnetization and b) the Binder cumulant are obtained by the quantum-inspired classical algorithm discussed in the main text. In this algorithm the Loschmidt echos $G_\psi(t)$ are numerically simulated and could alternatively be obtained from a quantum simulator; c.f. blue box in Fig.~\ref{fig:1}c). Error bars are estimated from a Jackknife binning analysis.
} 
 \label{fig:3}
 \end{figure}

\emph{\textbf{Quantum-inspired classical algorithm.---}}The algorithm described above can also be used as a purely classical method, requiring an exact method for calculating the Loschmidt echos $G_\psi(t)$ up to short times. We use the Time-Dependent Variational Principle (TDVP) for MPS~\cite{Haegeman2011,Haegeman2016}, which can be readily applied to systems with long-range interactions. In order to reach the required times, a relatively small bond dimension $\chi=15$ is sufficient (see supplement for convergence~\cite{supp}). 

In Fig.~\ref{fig:3} we show the observables obtained from the algorithm for system sizes $L=8-64$ and compare them to exact results from matrix-product operator based imaginary time evolution~\cite{Zaletel2015} of a purified MPS~\cite{Verstraete2004}. We find excellent agreement even in the vicinity of the phase transition. To simulate the algorithm, we used a timestep $J\Delta t=0.1-0.05$, approximately following the scaling $\Delta t \propto 1/\sqrt{L}$ noted above. The maximum time $J t_\mathrm{max}=2\,(1)$ for $L=8-32\,(64)$ is small. We chose a filter width $\delta/J=2\,(4)$ for $L=8-32\,(64)$, a cut-off $p_\mathrm{cut}=10^{-6}$ and averaged over $10-19$ independent MC runs with $N_\mathrm{MC}=6000, 10000, 2000, 4000$ iterations each for $L=8,16,32,64$, where the first $1000$ iterations were discarded as burn-in of the Markov chain. 

\emph{\textbf{Measurement of Loschmidt echo with trapped ions.---}}While we have now shown that the algorithm can detect the phase transition in the LTFIM once the Loschmidt echos $G_\psi(t)$ are known, we have yet to show how $G_\psi(t)$ can be measured in a trapped ion quantum simulator [step (2b) in the algorithm above]. To this end, consider the polar decomposition of $G_\psi(t)=re^{i\varphi}$. The absolute value $r$ is given by the probability of measuring $\ket{\psi}$ after time evolving $\ket{\psi}$ for a time $t$, i.e., $r^2=|\braket{\psi|e^{-i\hat Ht}|\psi}|^2$, which has previously been measured in trapped ions~\cite{Jurcevic2017}. The phase $\varphi$ can be obtained from a Ramsey-type experiment by interfering a state which evolves under $\hat H$ with one that does not. To engineer such a state, we introduce a shelving state $\ket{s}$ which does not couple to the qubit levels under $\hat H$, such as one of the D$_{5/2}$ Zeeman sublevels in ${}^{40}$Ca$^{+}$~\cite{Jurcevic2014} or the ${}^2$D$_{5/2}$ state in ${}^{171}$Yb$^{+}$~\cite{Edmunds2021}. This allows us to obtain the phase $\varphi$ by rotating into a superposition between $\ket{\psi}$ and $\ket{s\cdots s}$, evolving it in time, rotating back and measuring the return probability to $\ket{\psi}$ (see supplement~\cite{supp} and Ref.~\onlinecite{Lu2021}). However, this superposition is a GHZ-type state, which is in general difficult to prepare. 
 
\begin{figure}
\centering
\includegraphics[width=\columnwidth]{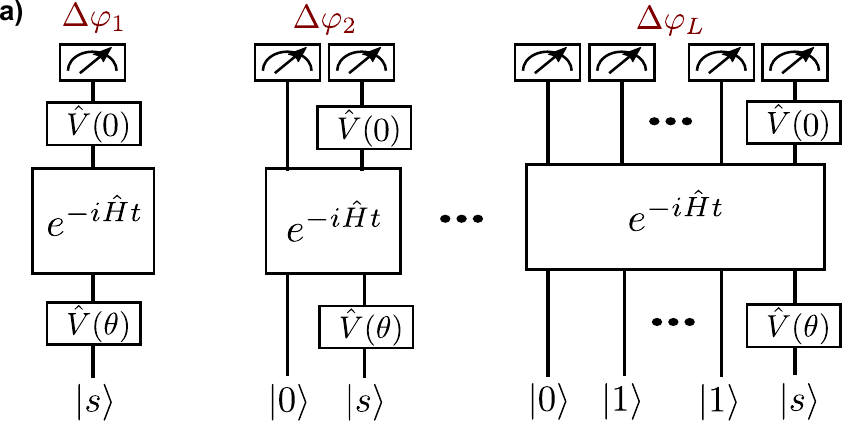}
\par
\includegraphics[width=\columnwidth]{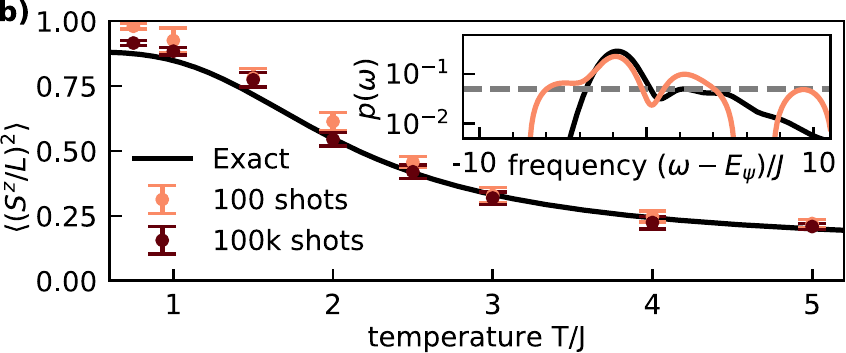}
\caption{\textbf{Ramsey protocol and robustness.} a) Pseudo-circuit diagram of the Ramsey protocol for measuring the phase. For $j=0$, $j=1$, etc. there are $L-1$, $L-2$, etc. additional ions in the $\ket{s}$ state which we do not display for simplicity as the time evolution acts trivially as an identity on these states. b) Results of a full simulation of the Ramsey protocol for a finite number of measurement repetitions. Inset: work distribution of totally polarized state, showing the influence of the noise. Grey dashed line illustrates $p_\mathrm{cut}$. We used $10^4$ MC iterations, $\delta=1$, $L=10$, $Jt_\mathrm{max}=4$, $J\Delta t=0.1$, cut $p_\mathrm{cut}=5 \times 10^{-2}$ (100 shots), $8 \times 10^{-4}$ (100k shots).}
 \label{fig:4}
 \end{figure}

To avoid the creation of a GHZ state, we can instead use a sequence of Ramsey experiments, akin to the sequential protocol proposed in Ref.~\onlinecite{Lu2021}: 
For each $j\in[0,L-1]$, we consider the state $\ket{\psi_j}$, in which the $j$ leftmost ions are in the qubit state corresponding to $\ket{\psi}$ and the rest are in the shelving state $\ket{s}$; c.f. Fig.~\ref{fig:4}. The phase difference $\Delta \varphi_{j+1} = \varphi_{j+1} - \varphi_j$ of the Loschmidt echos of two states is then obtained through the following Ramsey experiments:\\ 
(1) For a set of phases $\theta$:\\
\hspace*{10pt}(a) Prepare the state $\ket{\psi_j}$.\\
\hspace*{10pt}(b) Act with a single ion operation $\hat V_{j+1}(\theta)$ on ion $j+1$:\\
\hspace*{30pt}$\hat V_{j+1}(\theta)\ket{\psi_j}=\frac{1}{\sqrt{2}}(\ket{\psi_j}+e^{i\theta}\ket{\psi_{j+1}})$.\\
\hspace*{10pt}(c) Evolve with $\hat H$ for time $t$.\\
\hspace*{10pt}(d) Act with $\hat V_{j+1}^\dagger(0)$ on ion $j+1$.\\
\hspace*{10pt}(e) Measure the return probability $M(\theta)$ to state $\ket{\psi_j}$:
\begin{equation}
    M(\theta)=\frac{1}{4}\left( r_j^2+r_{j+1}^2+2r_{j+1}r_j\cos(\theta+\Delta \varphi_{j+1})\right).
    \label{eq:Mtheta}
\end{equation}
(2) Fit the measured $M(\theta)$ to Eq.~\eqref{eq:Mtheta}, to obtain the phase difference $\Delta \varphi_{j+1}$. We found this fit to perform best if $r_{j}$ and $r_{j+1}$ are also measured, such that $\Delta \varphi_{j+1}$ is the only unknown.\\
(3) Add up all $L$ phase differences to find $\varphi$. To see this, we use that $\varphi_0=0$ because $G_{\ket{s\cdots s}}=1$, which gives
\begin{equation}
    \varphi=\sum_{j=0}^{L-1} \Delta \varphi_{j+1}.
    \label{eq:varphi}
\end{equation}
We have therefore found an algorithm which determines the phase of $G_\psi(t)$ using $O(L)$ single-ion Ramsey experiments, that can be implemented directly on current experiments.

\emph{\textbf{Robustness.---}} In order to demonstrate the robustness of the algorithm, we consider several sources of error in the measurement of $G_\psi(t)$ in the following. In an experiment, $M(\theta)$ and $r$ can only be determined up to a precision $\sim 1/\sqrt{N_s}$ due to quantum projection noise~\cite{Itano1993} introduced by measuring $N_s$ times. In order to show that this algorithm works even for a finite $N_s$, we simulated the above Ramsey protocol and show results in Fig.~\ref{fig:4}b). We used $N_s$ repetitions for $r_j$ as well as $N_s/4$ repetitions for $M(\theta)$ for four equally spaced values of $\theta\in[0,\pi]$, such that $2N_sL$ measurements are performed per time point. The algorithm therefore requires $N_{\mathrm{MC}} (t_\mathrm{max}/\Delta t) 2 L N_s$ measurements, which reduces to $N_{\mathrm{MC}} (t_\mathrm{max}/\Delta t) N_s$ when using the GHZ protocol. The noise leads to errors when $p_\psi(\omega)$ is small, which we remove by using a cut $p_\mathrm{cut}\propto 1/\sqrt{N_s}$; c.f. Fig.~\ref{fig:4}b) inset. We find good results even for a small number of measurements. At low temperatures, more measurements are needed as small values of $p_\psi(\omega)$ for large negative $\omega$ become more important due to the factor $e^{-\omega/T}$ in Eq.~\eqref{eq:p_psi}. 

Another source of imprecision are errors in state preparation and measurement (SPAM). The leading contribution is given by uncorrelated single-qubit state assignment errors with probability $p$. The return probability $|G_\psi(t)|^2$ then has an error of $1-(1-p)^L$. Current trapped-ion devices have $p\approx 10^{-3}$ (see e.g. Ref.~\onlinecite{Joshi2022}), such that the error on $|G_\psi(t)|^2$ is $5\%$ for $L=50$. Uncorrelated SPAM errors can be corrected by multiplying the measured probability distribution with the  tensor product of the inverse single qubit measurement error matrices, see e.g.~\cite{Zheng2017_3,Satzinger2022}.

The effect of dephasing noise may be modelled by an exponential decay of the Loschmidt amplitude $G_\psi(t)\rightarrow G_\psi(t)\exp(-\gamma L t)$. The resulting $1/\omega^2$ tails in the work distribution lead to errors at large positive or negative frequencies. Similar to measurement noise, this error can in principle be mitigated by using a suitably small cut-off. If $\gamma$ is known, multiplying the measured signal with $\exp(\gamma L t)$ might lead to an even more efficient removal of decoherence effects. However, we expect decoherence effects not to be very strong as the time scales $Jt\sim 1-2$ are small compared to those routinely employed in trapped-ion quantum simulators. 

\emph{\textbf{Discussion \& Outlook.---}}To summarize, we have applied a protocol to probe finite temperature observables in analogue quantum simulators of spin systems. We benchmarked the protocol by studying a thermal phase transition in the transverse-field Ising mode with long-range interactions as realized with trapped-ions. 
This algorithm is well suited to current noisy devices as only short times are needed, independent of system size. Due to the importance sampling, the main bottleneck is the number of measurements. Current trapped ion simulators employing ${}^{40}$Ca$^{+}$ (${}^{171}$Yb$^{+}$) have a shot time on the order of $100$ms ($10$ms), such that $N_\mathrm{MC}=1000$ samples seem to be realistically achievable.

The algorithm can be directly employed in other platforms. Rydberg atoms in tweezers have access to efficient GHZ state preparation~\cite{Verresen2021}, shelving states as well as a high measurement repetition rate. They could probe 2D or 3D interacting Ising models in frustrated lattices, for which classical algorithms are plagued by the sign problem, thus providing a route to obtain quantum advantage~\cite{Troyer2005}. In these cases, finding more efficient update rules for the importance sampling, e.g. by using the quantum simulator for the proposal step~\cite{Layden2022} might be advantageous. Moreover, the free energy of the system can in principle be evaluated via $F=-T \ln{Z}$ using for example multi-canonical sampling strategies~\cite{Lyubartsev1992}. The range of models accessible in analogue simulators can be extended by applying prethermalization (e.g. the XY model in trapped ions~\cite{Neyenhuis2017}), or Floquet engineering (e.g. Heisenberg models~\cite{birnkammer2020,Geier2021}).

While our primary motivation was the application to quantum simulators, this algorithm can also be used to study previously inaccessible regimes with MPS methods due to the low requirement on the maximum time and hence the entanglement.

\textbf{Note added.---}During the completion of this manuscript, a work appeared proposing this algorithm as a quantum-inspired classical algorithm for MPS~\cite{Yang2022}.

\textbf{Acknowledgments.---}We thank Mari-Carmen Ba\~nuls, Ignacio Cirac, Henrik Dreyer, Kevin Hémery, Markus Heyl, Mohsin Iqbal, Manoj Joshi, Michael Labenbacher, Sirui Lu, Ramil Nigmatullin, Christian Roos, and Yilun Yang for discussions. 
We acknowledge support from the Deutsche Forschungsgemeinschaft (DFG, German Research Foundation) under Germany’s Excellence Strategy--EXC--2111--390814868, TRR80 and DFG grants No. KN1254/1-2, KN1254/2-1, the European Research Council (ERC) under the European Union’s Horizon 2020 research and innovation programme (grant agreement No. 851161), the Munich Quantum Valley, which is supported by the Bavarian state government with funds from the Hightech Agenda Bayern Plus, the German Federal Ministry of Education and Research (BMBF) through the funded project EQUAHUMO within the funding program quantum technologies - from basic research to market, as well as the NSF through a grant for the Institute for Theoretical Atomic, Molecular, and Optical Physics at Harvard University and the Smithsonian Astrophysical Observatory. Numerical simulations were performed using the TeNPy Library (version 0.9.0)~\cite{Hauschild2018} and QuSpin~\cite{Weinberg2017}.

{\textbf{Data and materials availability.---}}Data analysis and simulation codes are available on Zenodo upon reasonable request~\cite{Zenodo}.

\bibliography{literature.bib}

\appendix
 
\begin{onecolumngrid}

\section{Variance of $p_\psi(\omega)$}
The variance of $p_\psi(\omega)$ is defined as
\begin{equation}
    (\Delta p_\psi(\omega))^2 = \int d\omega \omega^2 p_\psi(\omega) - \left( \int d\omega \omega p_\psi(\omega) \right)^2.
\end{equation}
Using that 
\begin{align}
    p_\psi(\omega)&=\int dt e^{i\omega t} \braket{\psi|e^{-i\hat Ht}|\psi}\\
    &= \sum_n |\braket{\psi|n}|^2 \delta(\omega-E_n)
\end{align}
We find
\begin{equation}
    \int d\omega \omega^n p_\psi(\omega)=\braket{\psi|(\hat H)^n|\psi}
    \label{eq:powers}
\end{equation}
and therefore $(\Delta p_\psi(\omega))^2=\braket{\psi|(\hat H)^2|\psi}-(\braket{\psi|\hat H|\psi})^2$. Specializing to the case in the main text, i.e., z-product states and the long-range transverse field Ising model, we find for the standard deviation
\begin{equation}
    \Delta p_\psi(\omega)=g\sqrt{L},
\end{equation}
explicitly showing the $\sqrt{L}$ scaling discussed in the main text.

\section{Observables from the Jarzynski equality}
Here we show how to get from Eq.~(4) to Eq.~(5) in the main text, in particular showing how Eq.~(5) generalizes to arbitrary observables.
To do so, we start from Eq.~(4) and insert the Jarzynski equality in the form $Z=\int d\omega e^{-\omega/T} \sum_\psi p_\psi(\omega)$. We then shift $\hat H\rightarrow \hat H+h\hat O$ in the evaluation of $p_\psi(\omega)$ (c.f. Eq.~(3)). We find
\begin{align}
\braket{\hat O}_T&=-\frac{T}{Z}\frac{dZ}{dh}\bigg|_{h=0} \\&= -\frac{T}{Z} \int d\omega e^{-\omega/T} \sum_\psi\int \frac{dt}{2\pi} (-it) \braket{\psi|\hat O e^{-i\hat H t}|\psi} e^{i\omega t}.
\end{align}
By using $\int d\omega e^{-\omega/T} e^{i\omega t}=2\pi \delta(t+i/T)$, we get
\begin{equation}
   \braket{\hat O}_T = \frac{\sum_\psi p_{\psi,O}(T)}{\sum_\psi p_\psi(T)},
   \label{eq:arbobs}
\end{equation}
where 
\begin{equation}
    p_{\psi,O}(T)=\int d\omega e^{-\omega/T} \int \frac{dt}{2\pi}  e^{i\omega t} \bra{\psi} \hat O e^{-i\hat H t}\ket{\psi}.
\end{equation}
The quantity $G_{\psi,O}(t)=\bra{\psi} \hat O e^{-i\hat H t}\ket{\psi}$ needs to be measured separately to $G_\psi(t)$ by acting with $\hat O$ onto the state after time evolution. We can rewrite this equation  into a form useful for importance sampling by inserting a unity $1=p_\psi(T)/p_\psi(T)$,
\begin{equation}
   \braket{\hat O}_T = \frac{\sum_\psi  p_\psi(T) \frac{p_{\psi,O}(T)}{ p_\psi(T) }}{\sum_\psi p_\psi(T)}.
\label{eq:observablesapp}
\end{equation}
  Hence, we sample over the states $\ket{\psi}$, where the acceptance probability of a new state $\ket{\psi'}$ is given by $p_{\psi'}(T)/p_\psi(T)$. To evaluate the observable, we average over $\frac{p_{\psi,O}(T)}{ p_\psi(T) }$.  We can also rewrite Eq.~\eqref{eq:arbobs} slightly differently by inserting a unity by $1=p_\psi(\omega)/p_\psi(\omega)$ in the numerator,
\begin{equation}
    \braket{\hat O}_T = \frac{\sum_\psi \int d\omega e^{-\omega/T} p_\psi(\omega) \frac{p_{\psi,O}(\omega)}{p_\psi(\omega)}}{\sum_\psi \int d\omega e^{-\omega/T} p_\psi(\omega)},
\end{equation}
where $p_{\psi,O}(\omega)=\int \frac{dt}{2\pi}  e^{i\omega t} \bra{\psi} \hat O e^{-i\hat H t}\ket{\psi}$. In this case, we need to sample over both the states $\ket{\psi}$ as well as frequencies $\omega$, with the acceptance probability of a new state and frequency given by $e^{-(\omega'-\omega)/T} \frac{p_{\psi'}(\omega')}{p_\psi(\omega)}$. Observables are evaluated by averaging over $\frac{p_{\psi,O}(\omega)}{p_\psi(\omega)}$ in the end. This is equivalent to Eq.~(38) in Ref.~\onlinecite{Lu2021}.

Finally, for the case $\hat O\ket{\psi}=O_\psi\ket{\psi}$ considered in the main text, we find
\begin{equation}
    p_{\psi,O}(T) = O_\psi p_\psi(T) ,
\end{equation}
such that Eq.~\eqref{eq:observablesapp} reduces to Eq.~(5) of the main text.

The measurement of $p_{\psi,O}(T)$ can be avoided if powers of the Hamiltonian are to be evaluated. By using that 
\begin{equation}
    \braket{(\hat H)^n}_T= (-1)^n \frac{d^nZ}{d((1/T)^{n})}
\end{equation}
and inserting the representation of $Z$ in terms of the work distribution, we get
\begin{equation}
    \braket{(\hat H)^n}_T=\frac{\sum_\psi p_\psi(T) \frac{\tilde p_\psi(T)}{p_\psi(T)}}{\sum_\psi p_\psi(T)},
\end{equation}
with 
\begin{equation}
    \tilde p_\psi(T)=\int d\omega \omega^n e^{- \omega/T} p_\psi(\omega).
\end{equation}
Hence, only $p_\psi(\omega)$ needs to be measured to measure any power of the Hamiltonian. In particular, this way the specific heat $C_V=(\braket{\hat H^2}_T-\braket{\hat H}_T^2)/(LT^2)$ can be directly evaluated.

\section{Convergence of MPS}

\begin{figure}
    \centering
    \includegraphics{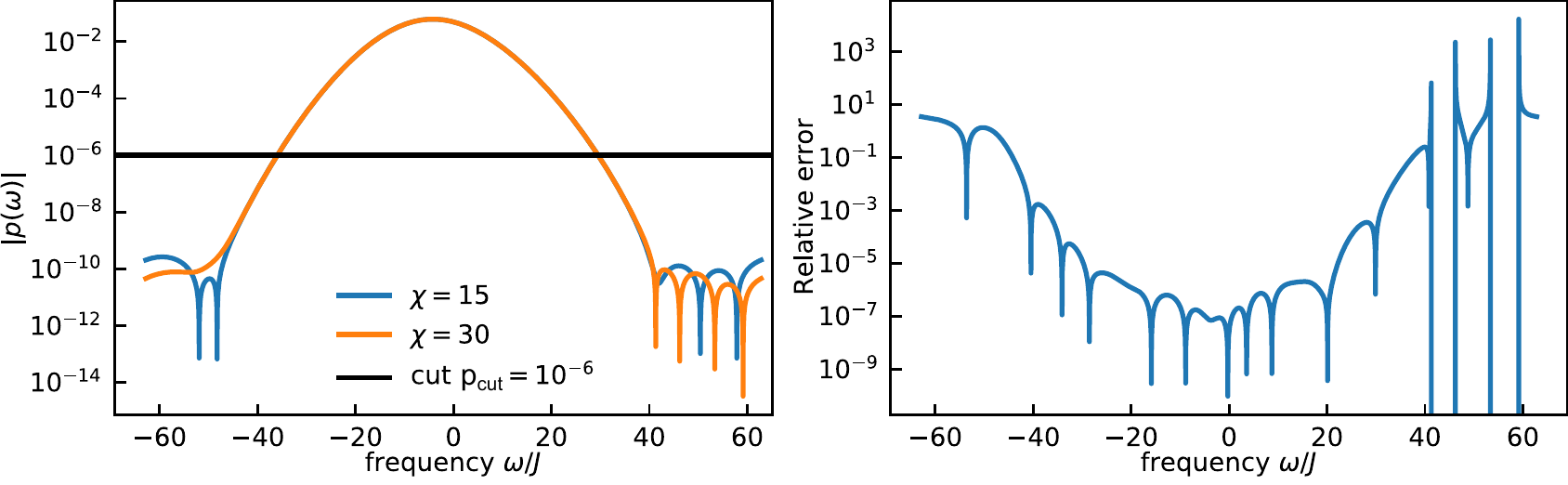}
    \caption{\textbf{Convergence with bond dimension.} Work distribution evaluated for a state with zero magnetization, system size $L=36$, $\delta/J=2$, $\Delta t=0.05$, $Jt_\mathrm{max}=2$. For all relevant frequencies, i.e., those with $p(\omega)>p_\mathrm{cut}$ (black, solid line), the relative error is smaller than $10^{-4}$. }
    \label{fig:Bonddim}
\end{figure}

In Fig.~\ref{fig:Bonddim} we show the work distribution for two values of the bond dimension, showing that small bond dimension $\chi=15$ indeed suffices to capture the dynamics. This is due to two effects: First, we only evolve to short times. Secondly, even within those short times, the later times do not contribute much to the low frequency behaviour of the work distribution because $G_\psi(t)$ has already decayed to small values for most $\ket{\psi}$. Even for states that do not decay quickly, such as the completely polarized state shown in Fig.~2a) of the main text, the latter fact is enforced by the filter, which suppresses late time contributions.

\section{Measurement of $G_\psi(t) = \bra{\psi}e^{-iHt}\ket{\psi}$ in analogue quantum simulators}

Here, we provide further details on the protocol to measure $G_\psi(t)$. As in Ramsey interferometry, a stationary path is compared to one which has evolved. 
We consider a system Hamiltonian $\hat H$ which acts on the states $\lbrace\ket{0},\ket{1}\rbrace$, where $\ket{\psi}$ is an arbitrary state in that basis. In addition, we introduce a shelving state $\ket{s}$ which does not evolve in time, $e^{-i\hat H t} \ket{s\cdots s}=\ket{s\cdots s}$.

The basic principle of the protocol is understood easiest when using a GHZ state---the single ion Ramsey protocol, which is easier to implement with quantum simulators, then follows.

\subsection{GHZ state protocol}
This protocol, similar to the ``cat state'' protocol in Ref.~\onlinecite{Lu2021}, uses the GHZ state
\begin{equation}
    \ket{\psi_\mathrm{GHZ}}=\frac{1}{\sqrt{2}} \left(\ket{0\cdots 0}+ \ket{1\cdots 1} \right)
\end{equation}
as a resource. The protocol proceeds as follows. First, transfer $0\rightarrow s$. Then, apply single qubit flips in $0-1$ subspace to rotate $\ket{1\cdots 1}$ into the target product state $\ket{\psi}$. Do this in a way so that a total phase $\theta$ is accumulated (e.g. by acting with a phase gate on one of the qubits), such that the system is now in
\begin{equation}
    \ket{\tilde \psi_\mathrm{GHZ} (\theta)} = \frac{1}{\sqrt{2}} \left(\ket{s\cdots s}+ e^{i\theta} \ket{\psi} \right).
\end{equation}
We define $\hat W(\theta)$ as the operation that takes $\ket{0\cdots0}$ to $\ket{\tilde \psi_\mathrm{GHZ}}$, i.e. $\ket{\tilde \psi_\mathrm{GHZ}}=\hat W(\theta)\ket{0\cdots0}$. After having prepared $\ket{\tilde \psi_\mathrm{GHZ} (\theta)}$, we evolve this state and apply $\hat W^\dagger(0)$. Finally, we measure the probability to be in $\ket{0\cdots 0}$, given by
\begin{align}
    N(\theta)&=|\bra{0\cdots 0} \hat W^\dagger(0) e^{-i\hat Ht } \ket{\tilde \psi_\mathrm{GHZ}}|^2\\
    &=\frac{1}{4} \left( r^2 +1+2r\cos(\theta+\varphi)\right),
\end{align}
where we inserted $G_\psi(t)=\braket{\psi|e^{-i\hat Ht}|\psi}=re^{i\varphi}$. By measuring $N(\theta)$ for several values of $\theta$, we can obtain $\varphi$ by fitting a cosine. What we have effectively done is a Ramsey experiment, where we used the fact that the state $\ket{s\cdots s}$ does not evolve in time. 

This protocol has the obvious disadvantage of needing a GHZ state, which is in general hard to prepare. We alleviate this in the following by replacing this protocol with a series of single ion Ramsey experiments.

\subsection{Single ion Ramsey protocol}
The single ion Ramsey protocol is explained in the main text. Here, we give some details on how to arrive at Eq.~(8). The return probability to $M(\theta)$ after step (1e) is given by
\begin{align}
    M(\theta)&=|\bra{\psi_j} \hat V^\dagger_{j+1}(0) e^{-i\hat H t} \hat V_{j+1}(\theta) \ket{\psi_j}|^2\\
    &= \frac{1}{4}|G_{\psi_j}(t)+e^{i\theta}G_{\psi_{j+1}}(t)|^2,
\end{align}
where we used that $\hat H$ cannot couple the states $\ket{\psi_j}$ and $\ket{\psi_{j+1}}$ because they differ in the number of qubits that are in the shelving state.
Inserting $G_{\psi_j(t)}=r_j e^{i\varphi_j}$, we get Eq.(8) of the main text: $M(\theta)=\frac{1}{4}\left( r_j^2+r_{j+1}^2+2r_{j+1}r_j\cos(\theta+\Delta \varphi_{j+1})\right)$.
It is advantageous to measure the amplitudes $r_j$ and $r_{j+1}$ occurring in this equation independently. Then only the phase of the cosine needs to be extracted from the fit by relating $y(\theta)=\cos(\theta+\varphi_{j+1}-\varphi_j)$ to
\begin{equation}
     y(\theta)= \frac{4M(\theta)-r_j^2-r_{j+1}^2}{2r_{j+1}r_j},
\end{equation}
which improves stability. 
This way, the phase difference $\varphi_{j+1}-\varphi_j$ is the only fitting parameter.

We note that the error on $y$ is inversely proportional to $r$. This may seem like a restriction when $r\rightarrow 0$, which is the case for large times, because then the error on $y$ will blow up. However, times for which $r$ is small do not contribute to $p(\omega)$. Hence, the increase of the error on $y$ for small $r$ is not important. We show explicitly in Fig.~4 of the main text that a small number of shots is sufficient.
\end{onecolumngrid}

\end{document}